\documentclass[nofootinbib,aps,12pt]{revtex4-1}
\usepackage{graphicx}
\usepackage{cancel}
\usepackage{textcomp}
\usepackage{amsmath}
\usepackage{bm}
\usepackage{epsfig}
\usepackage{color}
\usepackage{epstopdf}
\usepackage{dcolumn}
\usepackage{setspace}

\def\beq{\begin{equation}}
\def\eeq{\end{equation}}

\def\be{\begin{equation}}
\def\ee{\end{equation}}
\def\bea{\begin{eqnarray}}
\def\eea{\end{eqnarray}}

\begin{document}
\title{Stop-Neutralino Coannihilation in the Light of LHC
\vspace{4mm}
}
\author{M. Adeel Ajaib\footnote{email: adeel@udel.edu}}
\author{Tong Li\footnote{email: tli@udel.edu, corresponding author}}
\author{Qaisar Shafi\footnote{email: shafi@bartol.udel.edu}}
\address{
Bartol Research Institute, Department of Physics and Astronomy,
University of Delaware, Newark, Delaware 19716, USA
\vspace{3mm}
}

\begin{abstract}
We employ the ATLAS search results for events containing jets and large 
missing transverse momentum, corresponding to an integrated luminosity 
of 1 fb$^{-1}$, to investigate the constrained minimal supersymmetric model (CMSSM) with $b-\tau$ Yukawa coupling unification. In 
this scenario, one of the stops is the next-to-lightest supersymmetric 
particle (NLSP), which co-annihilates with the lightest (LSP) neutralino 
to yield the desired dark matter relic abundance.
The NLSP stop, here taken to be lighter than the top quark, is slightly ($\lesssim 20\%-30\%$) heavier than the LSP neutralino, and it primarily decays into the LSP and a charm quark. We find that the multi-jets and monojet ATLAS searches are sensitive to 
this scenario if the stop pair production is accompanied by a hard QCD jet.
The excluded limit for the NLSP stop mass from the ATLAS data can reach 160 GeV in the coannihilation region, with mass below 140 GeV essentially excluded. A significant region of the parameter space corresponding to large $m_0$ values, $8 \ {\rm TeV}\lesssim m_0\lesssim 16$ TeV, is excluded by our analysis. For LSP neutralino mass $\sim 100$ GeV, the LHC constraints in some cases on the spin-dependent (spin-independent) neutralino-nucleon cross section are significantly more stringent than the current and expected bounds from Xenon, CDMS and IceCube.
\end{abstract}
 \maketitle

\section{I\lowercase{ntroduction}}
Low scale supersymmetry, augmented by an unbroken R-parity, largely overcomes the gauge hierarchy problem encountered in the Standard Model (SM) and also provides a compelling cold dark matter candidate. In the mSUGRA/constrained minimal supersymmetric model (CMSSM)~\cite{Arnowitt:2006bb}, as well as in many other realistic models, the lightest neutralino (LSP) is stable~\cite{lsp} with a relic density that is compatible with the WMAP dark matter measurements~\cite{wmap}. However, the small annihilation cross section of a pure bino LSP with mass of around 100 GeV does not permit one to easily reproduce the required relic dark matter abundance~\cite{bino}. An interesting scenario which overcomes this conundrum is bino-NLSP stop coannihilation.
In this case the bino and the NLSP stop are quasi-degenerate in mass, such that the ensuing coannihilation processes in the early universe allow one to reproduce the desired dark matter relic density. Other related scenarios include bino-gluino coannihilation~\cite{binogluino} and bino-sbottom coannihilation~\cite{binosbottom}.

Besides the coannihilation scenario, there are other motivations for considering an NLSP stop. 
For instance, successful electroweak baryogenesis seems possible in the presence of a light stop and favors $M_{\tilde{t}_1}, M_h\lesssim 130$ GeV~\cite{wagner}. Also, the stop-top loop diagrams contribute to the parameter $m_{H_u}^2$, and naturalness in electroweak symmetry breaking in the MSSM consequently prefers an upper bound on the lighter stop mass of around a few hundred GeV~\cite{littlehierarchy}. The larger top Yukawa coupling typically yields
 third generation squark masses that are lighter than the first two generations~\cite{cohen}.


Our study here is inspired by a recent study~\cite{btau} which showed that SU(5) or SO(10) 
inspired $b-\tau$ Yukawa unification is compatible within the CMSSM 
framework with the WMAP dark matter bounds only if there exists NLSP 
stop - LSP neutralino coannihilation. The universal scalar parameter 
$m_0$ in this scenario turns out to be of order $5-20$ TeV, with the universal trilinear 
scalar coupling $A_0$ of comparable magnitude. The CMSSM parameter $\tan\beta\sim 35$ and 
$m_{1/2} \ll m_0$, where $m_{1/2}$ denotes the universal gaugino mass. 
Values of $m_0$ and $|A_0|$ of comparable magnitude have been discussed from a different perspective string inspired models in Ref.~\cite{daniel}. 

The search for NLSP stop, especially in the region of nearly degenerate stop and LSP neutralino masses, is challenging and has been implemented by both LEP and Tevatron~\cite{pdg,jose,cdf}. In this case, the stop two-body decay channels into a top quark and neutralino, or a bottom quark and chargino, and three-body decay channels $\tilde{t}_1\to W^+b\tilde{\chi}_1^0$, $\tilde{t}_1\to b\ell^+\tilde{\nu}$ and $\tilde{t}_1\to b\tilde{\ell}^+\nu$ are all kinematically forbidden. The loop-induced stop two-body decay into a charm quark and a neutralino is generally considered to overwhelm the four-body channel $\tilde{t}_1\to \ell^+\nu(q\bar{q}')b\tilde{\chi}_1^0$ and tends to be the dominant NLSP stop decay mode~\cite{kobayashi}. The most stringent mass limit on a light stop with decay into charm quark and LSP neutralino comes from the CDF search for events containing two jets and missing transverse energy, namely $M_{\tilde{t}_1}>180$ GeV~\cite{cdf}. However, the Tevatron is not sensitive to stop searches if the stop and LSP neutralino mass difference is below 40 GeV. Thus the Tevatron bound does not cover the coannihilation region above the LEP limit of $M_{\tilde{t}_1}\approx 100$ GeV~\cite{pdg}. 

Two alternative search methods have been investigated to detect a light stop. One of them takes advantage of the Majorana fermion feature of gluino and considers gluino pair production followed by gluino decay into on-shell stop and top quark~\cite{kraml,martinss}. The pair production of gluinos leads to events containing a pair of same-sign top quarks plus two same-sign stops. The benefit of this search is the anomalous same-sign dilepton signature arising from the same-sign top quarks leptonic decay, with negligible SM backgrounds. However, this method strongly depends on the production of relatively light gluino. From the well-known relation between gaugino masses at low energy in mSUGRA/CMSSM, namely $M_3:M_2:M_1\approx 6:2:1$, which follows from the assumption of universal gaugino masses at high scale, the gluino masses have to be above at least $600-700$ GeV for the coannihilation region of NLSP stop and LSP neutralino with $M_{\tilde{t}_1}> 100$ GeV. This leads to small production cross sections, and so it is safe to assume that this scenario could not produce a significant amount of gluino pair events with same-sign dileptons at this early stage of LHC and evades the current bound especially for large values of $m_0$~\cite{cmsss}. The other proposed method is to consider stop pair production associated with a hard jet~\cite{wagnerstop}. In the coannihilation region, there will be minimal hadronic activity associated with the stop decay and therefore this channel would effectively lead to events with a hard jet and large missing energy. Such a signature has been proposed to explore large extra dimensions~\cite{led}, search for relatively light gluinos at Tevatron~\cite{jgogo}, and nearly degenerate gaugino pair production~\cite{tao}.

Recently, ATLAS reported results on an inclusive search for new phenomena in an event sample with monojet and large missing transverse momentum in the final state corresponding to an integrated luminosity of 1 fb$^{-1}$~\cite{atlasmonojet}. Good agreement was observed between the number of events in the data and the SM predictions. The results can be translated into improved limits on the stop-neutralino coannihilation scenario in terms of stop pair production associated with a hard jet. Also, the ATLAS and CMS experiments at $\sqrt{s}=7$ TeV LHC have presented their results for events containing jets and missing transverse momentum for low-energy supersymmetry in 2011~\cite{atlasjets,cmsjets}. In mSUGRA/CMSSM with $\tan\beta=10, A_0=0$ and $\mu>0$, squarks and gluinos of equal mass are excluded for masses below 1 TeV or so. The lower limits on fundamental parameters of mSUGRA/CMSSM are $m_{1/2}\sim 500$ GeV and $m_0\sim 3$ TeV. However, the current ``low statistics'' with integrated luminosity up to 1 fb$^{-1}$ encourages searches involving cascades initiated by gluino and the first two generation squarks, and consequently the limits do not significantly depend on $\tan\beta$ and $A_0$ parameters in mSUGRA/CMSSM~\cite{nath1}. An analysis for scenarios with rich production of the third generation squarks induced by large values of $m_0$ and $A_0$ in mSUGRA/CMSSM is still lacking.

In this paper we study the LHC constraints on NLSP stop scenario in $b-\tau$ Yukawa unified mSUGRA/CMSSM using the monojet and multiple jets search results from the LHC. The paper is organized as follows. In section II we summarize the NLSP stop scenario with $b-\tau$ Yukawa unification in mSUGRA/CMSSM and LSP neutralino (essentially bino) dark matter (for more details, see Ref.~\cite{btau}). We also discuss the NLSP stop production modes and outline the selection cuts employed by the ATLAS collaboration. The results of the NLSP stop model constrained by the LHC data 
are presented in section III. Our conclusions are summarized in section IV.
 
\section{NLSP S\lowercase{top} \lowercase{in} $b-\tau$ Y\lowercase{ukawa} U\lowercase{nified} \lowercase{m}SUGRA/CMSSM \lowercase{and} ATLAS S\lowercase{election} R\lowercase{equirements}}
As mentioned earlier, the stop-bino coannihilation scenario requires the stop to be the NLSP in the sparticle spectrum, and to be nearly degenerate in mass with the bino LSP. The mass difference between the two should be~\cite{bino}
\begin{eqnarray}
{M_{\tilde{t}_1}-M_{\tilde{\chi}_1^0}\over M_{\tilde{\chi}_1^0}}\lesssim 20\%.
\label{coann}
\end{eqnarray}
This scenario can be realized in mSUGRA/CMSSM~\cite{drees,drees1,nathstop} because, compared to the other squarks, $M_{\tilde{t}_1}$ is reduced by contributions of the
large top quark Yukawa coupling to the relevant renormalization group equations, as well as by 
mixing between the $SU(2)$ doublet and singlet top squarks. To implement $y_b=y_\tau$ at $M_{GUT}$, sizable threshold correction to the bottom Yukawa coupling $\delta y_b$ is required~\cite{btau}
\begin{eqnarray}
\delta y_b^{\rm finite} \approx {\mu\over 4\pi^2}\left({g_3^2\over 3}{M_{\tilde{g}}\over \bar{M}_1^2}+
{y_t^2\over 8}{A_t\over \bar{M}_2^2}\right) \tan\beta,
\label{yb}
\end{eqnarray}
where $g_3$ is the strong gauge coupling, $M_{\tilde{g}}$ is the gluino mass, and $\bar{M}_1\approx (M_{\tilde{b}_1}+M_{\tilde{b}_2})/2$, $\bar{M}_2\approx (M_{\tilde{t}_2}+\mu)/2$. Also, the hierarchy $M_{\tilde{g}}\ll M_{\tilde{b}_1},M_{\tilde{b}_2}; M_{\tilde{t}_1}\ll \mu, M_{\tilde{t}_2}$ is assumed. For $\mu>0$, to get the correct (negative) threshold correction, the contribution from the chargino loop (the second term of Eq.~(\ref{yb})) should not only cancel the contribution from the gluino loop (the first term of Eq.~(\ref{yb})), but it also must provide the correct negative sign for $\delta y_b$. This, it turns out, is achieved only with large values of $m_0$ and $|A_0|$, with $\tan\beta\sim 35$.

The software package ISAJET 7.80~\cite{isajet} was employed in Ref.~\cite{btau} to scan over the relevant four parameters with $\mu>0$, as well as renormalization group evolution of gauge and Yukawa couplings and all soft parameters, and finally the computation of the physical masses of all particles. 
A large number of relevant phenomenological constraints such as $BR(b\to s\gamma)$~\cite{bsr}, $BR(B_u\to \tau\nu)$~\cite{bsr}, LEP II bound on the lightest Higgs and all the sparticle mass bounds~\cite{mass} were also implemented. Note that recently the CDF collaboration has reported an excess in the rare decay $B_s^0\to \mu^+\mu^-$ using 7 fb$^{-1}$ of integrated luminosity and the measured central value of $BR(B_s^0\to \mu^+\mu^-)$ is at least five times larger than the expected SM value~\cite{bscdf}. However, the combination of CMS and LHCb searches using 0.34 fb$^{-1}$ and 1.14 fb$^{-1}$ integrated luminosity has not confirmed this excess. Indeed, they provide a more stringent upper limit on the branching ratio, namely $BR(B_s^0\to \mu^+\mu^-)<1.08\times 10^{-8}$ at 95$\%$ confidence level~\cite{bslhc}. In our analysis we apply the upper limit from LHC.

The degree of Yukawa unification is quantified by the parameter $R$~\cite{r}  
\begin{eqnarray}
R\equiv {{\rm max}(y_b,y_\tau)\over {\rm min}(y_b,y_\tau)} \ .
\end{eqnarray}
In Ref.~\cite{btau}, $R$ is required to be $\leq 1.1$, so that  $b-\tau$ Yukawa unification holds at 10\% level or better. It was shown that, to get good $b-\tau$ Yukawa unification in mSUGRA/CMSSM, the parameters lie in the range of $5 \ {\rm TeV}\lesssim m_0\lesssim 20 \ {\rm TeV}, 35\lesssim \tan\beta \lesssim 40$, with $|A_0/m_0|\sim 2.3$ and $m_{1/2}\ll m_0$. The NLSP stop is quasi-degenerate in mass with LSP neutralino, and in our study it will be lighter than the top quark. The gluino is about 6 times heavier than the LSP neutralino, while the remaining sfermions all have much heavier masses, namely greater than 5 TeV for $M_{\tilde{t}_2}, M_{\tilde{b}_{1,2}}, M_{\tilde{\tau}_{1,2}}$ and 10 TeV for the first two family squarks.

Following the theoretical estimates in Refs.~\cite{kobayashi,drees,shih} and experimental search assumption~\cite{cdf}, we assume that the NLSP stop decays with 100\% branching fraction into a charm quark and neutralino LSP. Also, the total decay widths of the models we consider are of order $10^{-10}$ GeV, which guarantees that the stops promptly decay in the detector (the decay length is too short to be displaced).

For the NLSP stop scenario, the pure stop pair production is the leading source of stops. However, the small mass difference between the NLSP stop and LSP neutralino induces very soft charm jets and low missing energy from NLSP stop decay, which very likely evades the current LHC search bounds for multiple energetic jets. At best, only a tiny range of light NLSP stops could pass the relevant selection cuts because of the relatively large cross sections. It is therefore important to include the hard QCD emission at the matrix element level in order to provide a hard jet and large missing energy, and thus explore more stringent limits on the NLSP stop. In this scenario the heavier gluino essentially decays into on-shell stop plus top quark. 
The energetic particles from the top decay could compensate for the loss of events arising from the low cross section of heavy gluino production and NLSP stop decay followed by a soft jet.
 Based on these considerations, we generate hard scattering processes of gluino pair production and stop pair production together with the same processes with one extra jet at the matrix element level using Madgraph/Madevent~\cite{MG}
\begin{eqnarray}
pp\to \tilde{g}\tilde{g},\ \tilde{t}_1\tilde{t}_1^\ast,\ j\tilde{t}_1\tilde{t}_1^\ast,
\label{pro}
\end{eqnarray}
with the gluino decaying with 50\% branching ratio into $t\tilde{t}_1^\ast$ and $\bar{t}\tilde{t}_1$ each. We use Pythia to include decays and parton showering and hadronization~\cite{Pythia}, and PGS-4 to simulate the important detector effects with ATLAS-like parameters~\cite{PGS}. We take care to correctly match (without double-counting) between matrix element and showering generation of additional jets. In Madgraph/Madevent running we implement MLM matching with $P_T$-ordered showers and the shower-$K_T$ scheme with $Q_{cut}=100$ GeV as described in Ref.~\cite{matching}. The cross sections are normalized to the next-to-leading order output of Prospino 2.1~\cite{prospino}.

The ATLAS and CMS collaborations have reported data in terms of events containing jets and large missing transverse momentum in $\sqrt{s}=7$ TeV proton-proton collisions, corresponding to an integrated luminosity of 1 fb$^{-1}$~\cite{atlasjets,cmsjets}. Taking ATLAS as an example, the selected events are required to have a leading jet $p_T$ of at least 130 GeV and other multiple jets with $p_T$ greater than 40 GeV. The quantity $m_{eff}$, the scalar sum of $\cancel{E}_T$ and the transverse momenta of the highest $p_T$ jets, must be more than 1 TeV in most of the selection modes. Apparently, these selection requirements are too stringent for our case with nearly degenerate NLSP stop and LSP neutralino. In our case, most of the jets come from the NLSP stop decay and are forced to be kinematically extremely soft. Thus, most of the events with a given stop mass would be eliminated by the selection cuts. We therefore expect that the upper limit on the excluded stop mass for nearly degenerate NLSP stop and LSP neutralino scenario from the multi-jets search would be lower as compared with the bounds on the gluino and the first two family squarks, although the contribution of the additional jet in Eq.~(\ref{pro}) would improve the situation.

More importantly for this scenario, the ATLAS experiment has looked for monojet plus missing energy events with the same 1 fb$^{-1}$ integrated luminosity~\cite{atlasmonojet}. They searched for one extremely hard jet, large missing energy and nothing else. No excess above the SM background expectation was observed. With more strict selection cuts and more data, new lower bounds on non-SM cross sections are obtained that are roughly 5 times more stringent than those from the 2010 data. This analysis has been used to make constraints on large extra dimensions~\cite{atlasmonojet} and model-independent interactions of dark matter~\cite{tim}. 
In our case, this search would be sensitive to stop pair production associated with one hard jet, followed by stop decay into a soft jet and missing energy. This can be employed, as we show here, to find useful constraints on the NLSP stop scenario with nearly degenerate stop and LSP neutralino masses.

In the search for monojet plus large missing transverse momentum, the signal events are selected according to 3 different cut requirements, named ``LP'', ``HP'' and ``VHP''~\cite{atlasmonojet} as shown in Table~\ref{cuts1}. The LP (HP) selection requires a jet with $p_T>120$ GeV ($p_T>250$ GeV), $|\eta^{jet}|<2$ in the final state, and $\cancel{E}_T>120$ GeV ($\cancel{E}_T>220$ GeV). Events with a second leading jet $p_T$ above 30 GeV (60 GeV) in the region $|\eta|<4.5$ are rejected. For the HP selection, the $p_T$ of the third leading jet must be less than 30 GeV, and an additional requirement on the azimuthal separation $\Delta \phi(jet,\vec{p}_T^{miss})>0.5$ between the missing transverse momentum and the direction of the second leading jet is required. This cut is used to select events with the first and second jets going in roughly the same direction to reduce background from $j(W\to)\tau\nu$ as stated in Refs.~\cite{wagnerstop,ledexp}. The VHP selection is defined with the same requirements as in the HP region, but with thresholds on the leading jet $p_T$ and $\cancel{E}_T$ increased up to 350 GeV and 300 GeV, respectively. Also, events with charged leptons are also rejected. The 95$\%$ C.L. upper limits on effective cross section (cross section times acceptance) for non-SM processes for signal region LP, HP, VHP are also shown in the last row of Table~\ref{cuts1}. 
Following Ref.~\cite{martin} we apply $\sigma\times {\rm acceptance}>\sigma_{\rm exp}$ as the exclusion requirement for each model, where $\sigma$ is the relevant total cross section and the acceptance is the ratio of signal events after and before selection cuts which reflects the effects of experimental efficiency. Note that the upper limit on the second/third leading jet may remove some of our signal events once the hard leading jet is required. It is because the subleading jets in the roughly opposite direction may become correspondingly harder to balance the large $p_T$ of the leading jet, although the sum of neutralino momenta leads to a large amount of missing energy. Nevertheless, we still apply this cut as we follow the ATLAS limits for $\sigma_{\rm exp}$, and it does help to efficiently suppress the $W/Z+{\rm jets}$ and $t\bar{t}$ backgrounds.


\begin{table}[tb]
\begin{center}
\begin{tabular}[t]{|c|c|c|c|}
  \hline
 & LP & HP & VHP\\
  \hline
 Leading jet $p_T$ (GeV) & $>120$ & $>250$ & $>350$\\
  \hline 
 Second jets $p_T$ (GeV) & $<30$ & $<60$ & $<60$ \\
  \hline
 Third jets $p_T$ (GeV) & $-$ & $<30$ & $<30$ \\
  \hline
  $\Delta \phi(\vec{p}_T^{{\rm miss}}, j_{2})$ & $-$ & $>0.5$ & $>0.5$ \\
  \hline
  $\cancel{E}_T$ (GeV) & $>120$ & $>220$ & $>300$\\
  \hline
  ATLAS $\sigma_{{\rm exp}}$ (pb) & $1.7$ & $0.11$ & $0.035$\\
  \hline
\end{tabular}
\end{center}
\caption{Summary of selection cuts and 95$\%$ C.L. upper limits on the effective cross section for non-SM processes for signal region LP, HP and VHP containing final states with monojet and missing transverse momentum with 1 fb$^{-1}$ luminosity, following the ATLAS data analyses~\cite{atlasmonojet}.}
\label{cuts1}
\end{table}


\section{ATLAS C\lowercase{onstraints} \lowercase{on} NLSP S\lowercase{top} \lowercase{and} N\lowercase{eutralino} D\lowercase{ark} M\lowercase{atter}}


To study the LHC constraints on this class of models, we generate more than half a million models by scanning the parameter space.
From these, 3705 models pass the various experimental constraints listed in section II, and they have acceptable Yukawa unification ($R\leq 1.1$) and NLSP stop. Note that in our analysis we focus on 983 of them corresponding to $M_{\tilde{t}_1}<M_t=173.3$ GeV, which also happens to be favored by electroweak baryogenesis.

\begin{figure}[tb]
\begin{center}
\includegraphics[scale=1,width=8cm]{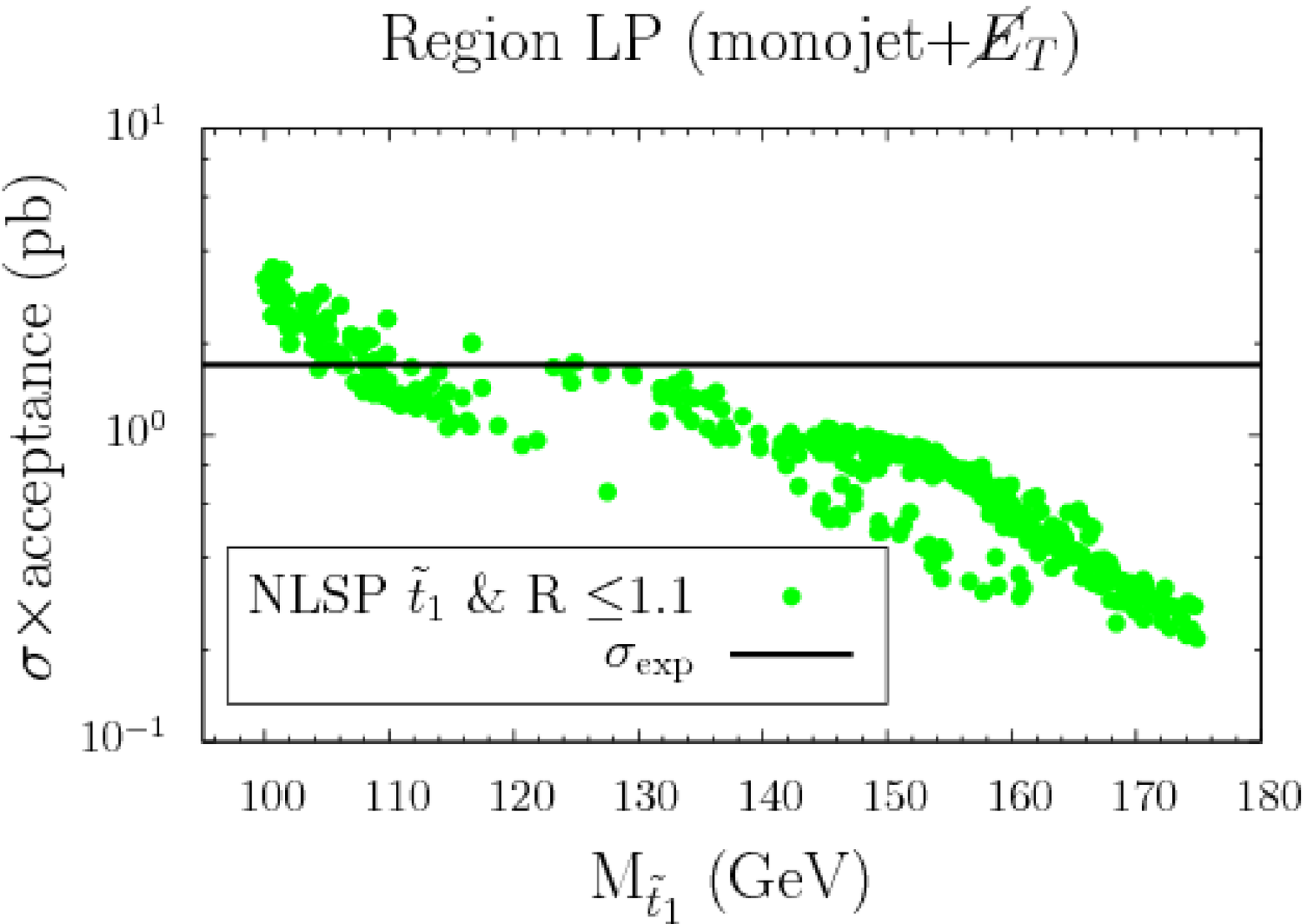}
\includegraphics[scale=1,width=8cm]{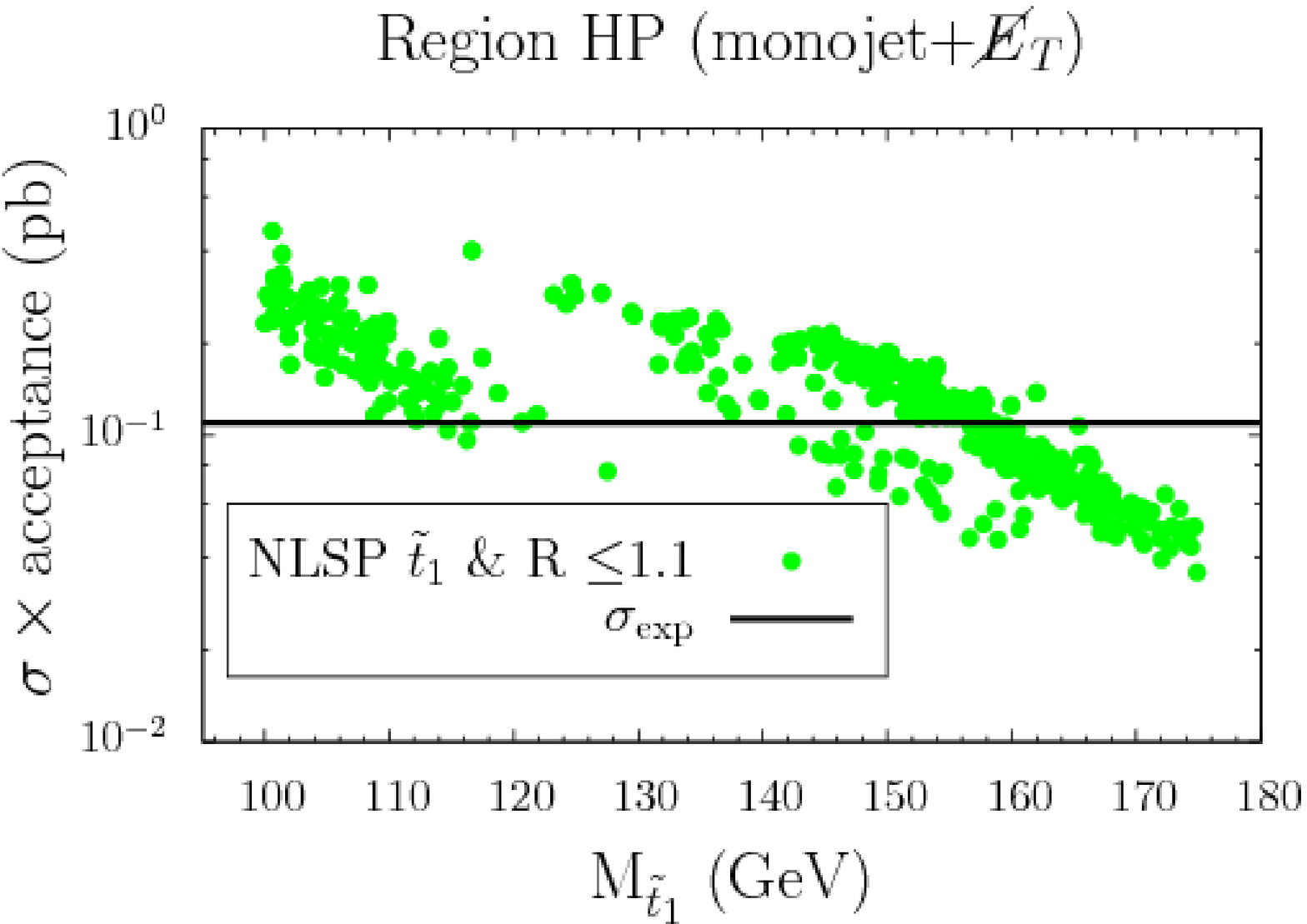}\\
\includegraphics[scale=1,width=8cm]{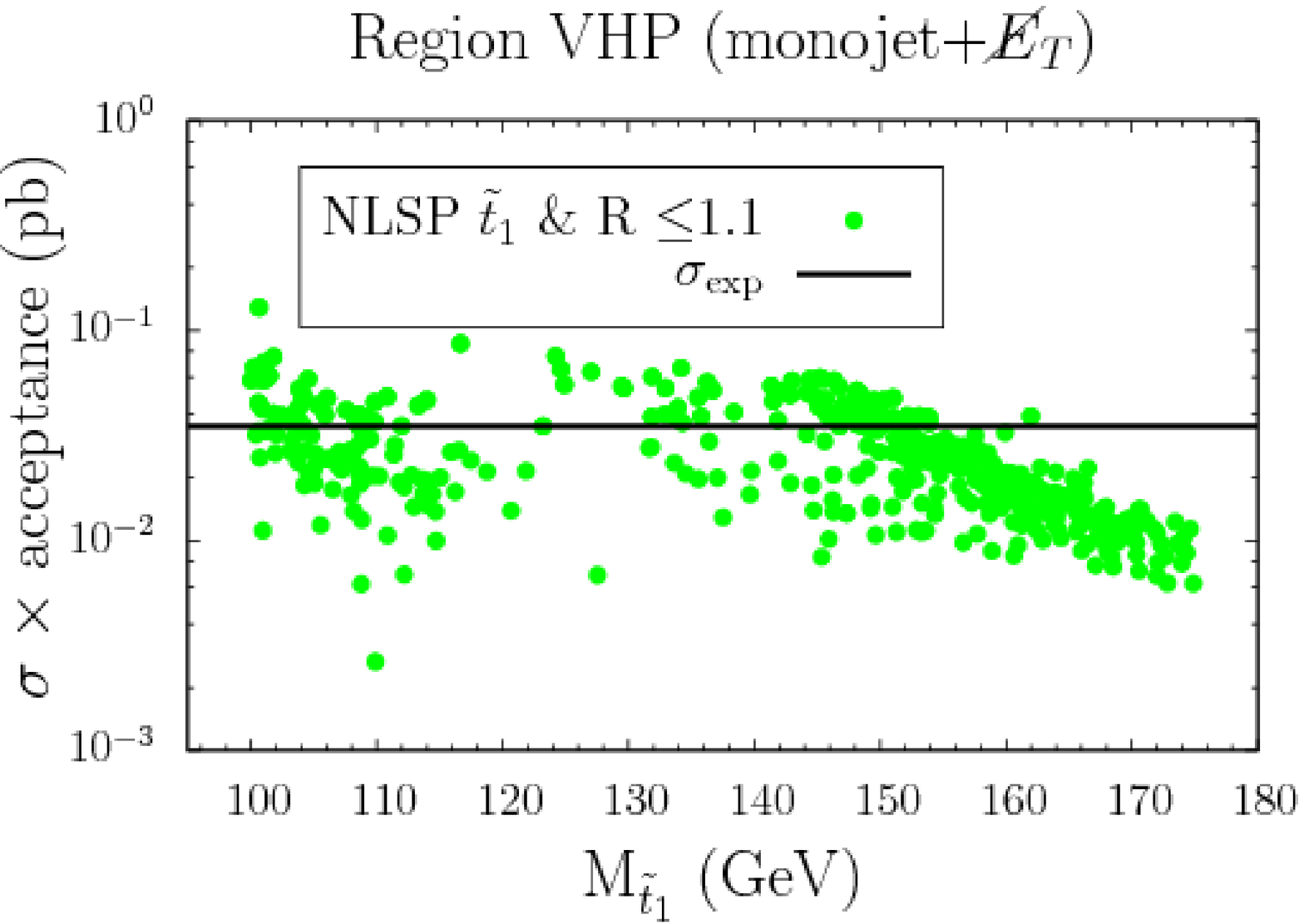}
\end{center}
\caption{$\sigma\times$acceptance vs. $M_{\tilde{t}_1}$ with horizontal line as the 95$\%$ C.L. upper limits on effective non-SM processes cross section for signal region LP (top left), HP (top right), VHP (bottom).
Green regions correspond to models with Yukawa unification ($R\leq 1.1$) and NLSP stop.} 
\label{a}
\end{figure}

In Fig.~\ref{a} we show $\sigma\times$acceptance vs. $M_{\tilde{t}_1}$ for the models with Yukawa unification and NLSP stop, after applying the requirements in the ATLAS monojet regions LP, HP and VHP. The constrained values of NLSP stop mass increase as the required $p_T$ of leading jet gets higher in the three different regions as shown in Table~\ref{cuts1}, namely $M_{\tilde{t}_1}\sim 110$ GeV for LP region, and $M_{\tilde{t}_1}\sim 160$ GeV for the HP and VHP regions. It is because the emitted jet recoils against the two associated stops in the transverse direction to the beams, and thus its $p_T$ is somewhat correlated with the relevant stop mass. Combining the exclusions from the three regions, an NLSP stop mass below 140 GeV is essentially excluded. 

\begin{figure}[tb]
\begin{center}
\includegraphics[trim = 5mm 75mm 5mm 65mm,scale=.7,clip]{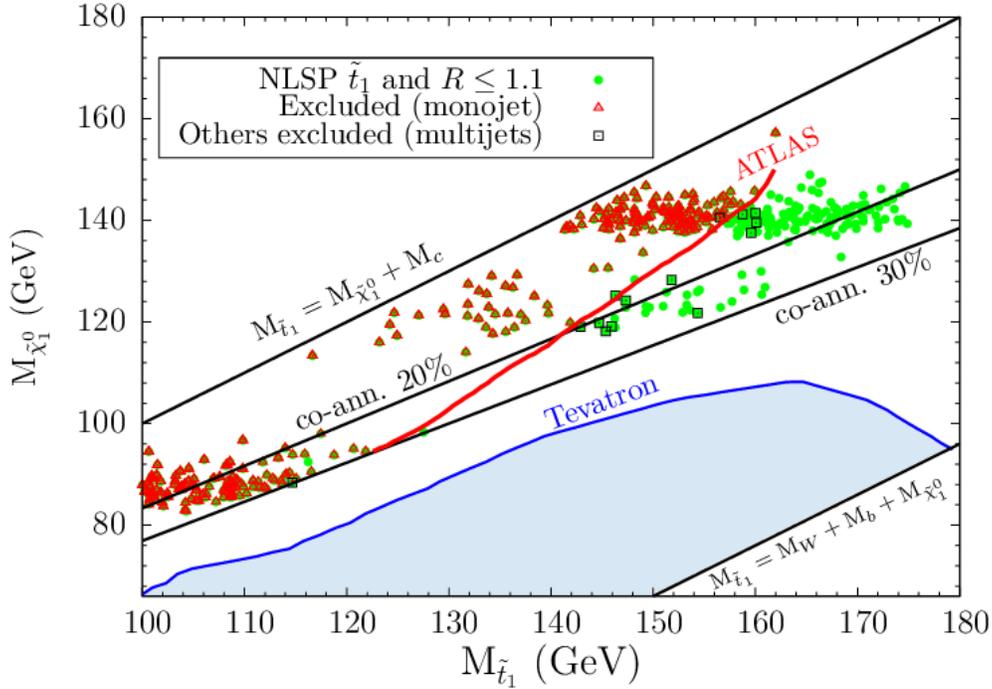}
\end{center}
\caption{$M_{\tilde{\chi}_1^0}$ vs. $M_{\tilde{t}_1}$ for models with Yukawa unification and NLSP stop (green circle), those excluded by ATLAS monojet regions (red triangle) and other excluded ones by ATLAS multi-jets regions (black box) in the framework of $b-\tau$ Yukawa unified mSUGRA/CMSSM. The kinematic limits and coannihilation bounds are also displayed. The blue region refers to the excluded region by Tevatron~\cite{cdf}.} 
\label{mstop}
\end{figure}

Fig.~\ref{mstop} shows the exclusion plot in the $M_{\tilde{\chi}_1^0}-M_{\tilde{t}_1}$ plane, with Yukawa unification and NLSP stop models (green circle). The top line corresponds to the kinematic bound of $\tilde{t}_1\to c\tilde{\chi}_1^0$ channel which is open below this line. The region below the bottom most straight line corresponds to the stop decay channel $\tilde{t}_1\to bW^+\tilde{\chi}_1^0$. In the region between these two lines, a stop decay into a charm quark and LSP neutralino is the unique channel for our study, since we assume that the 4-body channel is always highly suppressed. The coannihilation bounds from Eq.~(\ref{coann}) are also displayed in this plot. One can see that Tevatron bound does not cover the coannihilation region. 
However, the ATLAS monojet search does make additional inroads beyond the Tevatron, denoted by red triangles. For coannihilation region, it is more sensitive as the mass difference between the NLSP stop and LSP neutralino decreases. For the region with $M_{\tilde{t}_1}\gtrsim 140$ GeV the monojet search loses its capability when the mass difference is larger than 20 GeV because the charm jets from stop decay in this case become harder and cannot pass the $p_T$ selection requirement for the non-leading jets. 

Besides the monojet channels, we also apply the ATLAS multi-jets search requirements~\cite{atlasjets} and show the excluded models (but not by monojet search) with black box in Fig.~\ref{mstop}. The requirement of the additional jet and heavier gluino also provides events with hard jet(s) and large missing energy that pass the multiple energetic jets search cuts.
These excluded points gather in the region with $M_{\tilde{t}_1}-M_{\tilde{\chi}_1^0}\gtrsim 20$ GeV because of the induced relatively large $p_T$ of jets from stop decay. Based on these features we can clearly identify the region excluded by the LHC in Fig.~\ref{mstop}.

\begin{figure}[tb]
\begin{center}
\includegraphics[scale=1,width=11cm]{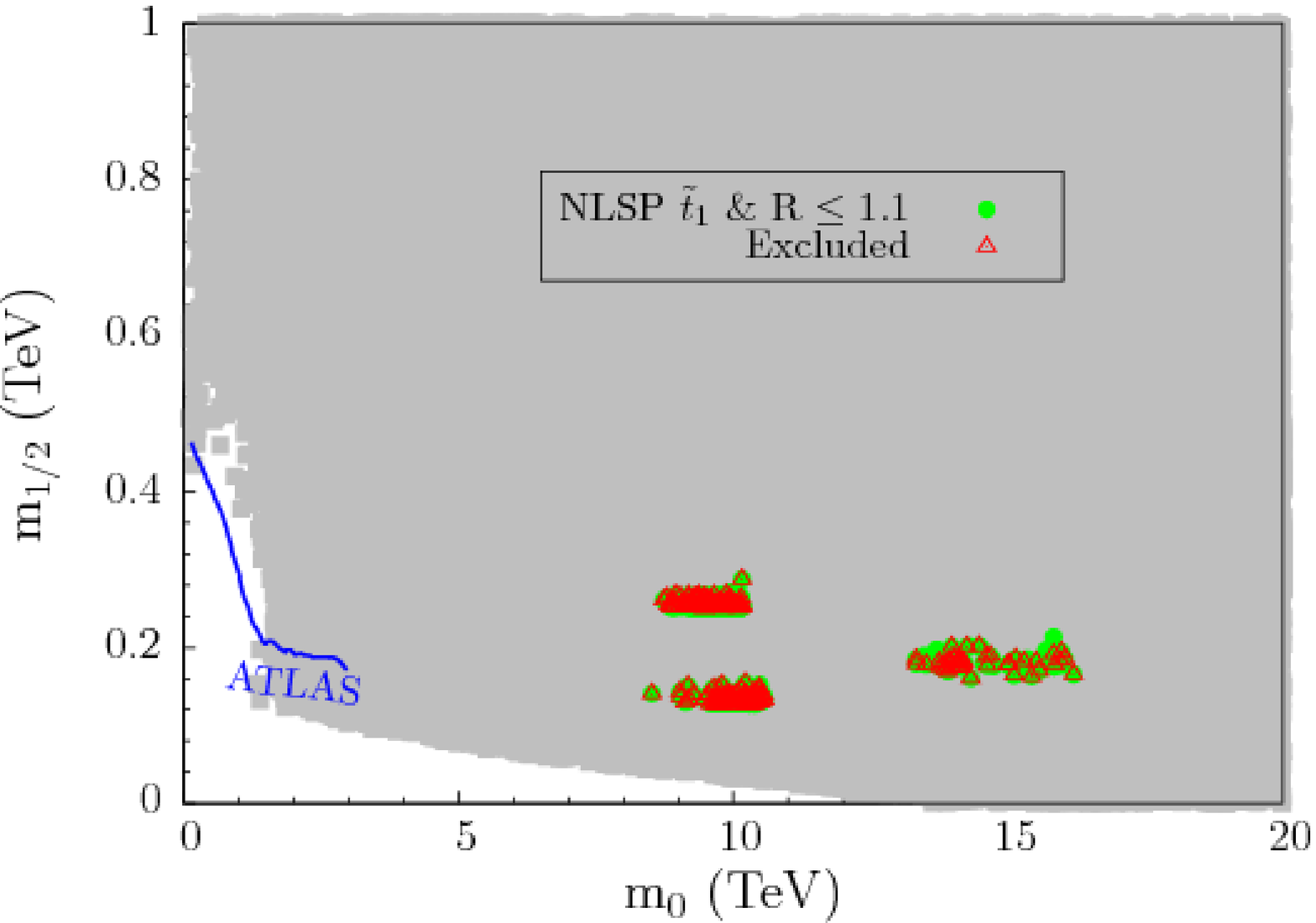}
\end{center}
\caption{$m_{1/2}$ vs. $m_0$ for models satisfying all low energy experiments (grey box), those with Yukawa unification and NLSP stop (green circle) and excluded ones by combined ATLAS monojet and multi-jets searches (red triangle) in the framework of $b-\tau$ Yukawa unified mSUGRA/CMSSM. The most stringent bound on this plane from ATLAS is also displayed~\cite{atlasjets}.} 
\label{mm}
\end{figure}

In Fig.~\ref{mm}, the excluded models are displayed in the $m_{1/2}-m_0$ plane with $\mu>0$ and varying $A_0$ and $\tan\beta$. We display all models that survive the low energy experiments listed in section II (grey color), models with good Yukawa unification and NLSP stop (green circle), and excluded models by the combined monojet and multi-jets searches (red triangle). One can see that the most stringent lower limit on $m_0$ is around 3 TeV for $\tan\beta=10, A_0=0, \mu>0$ case, from the LHC searches corresponding to comparable gluino and the first two family squarks masses~\cite{atlasjets}. Our models with good Yukawa unification and NLSP stop correspond to $m_0>8$ TeV and the above study on the production of NLSP stop approach much larger values of $m_0$, namely $8 \ {\rm TeV}<m_0<16$ TeV. A significant region of the parameter space is excluded.

\begin{figure}[tb]
\begin{center}
\includegraphics[scale=1,width=8cm]{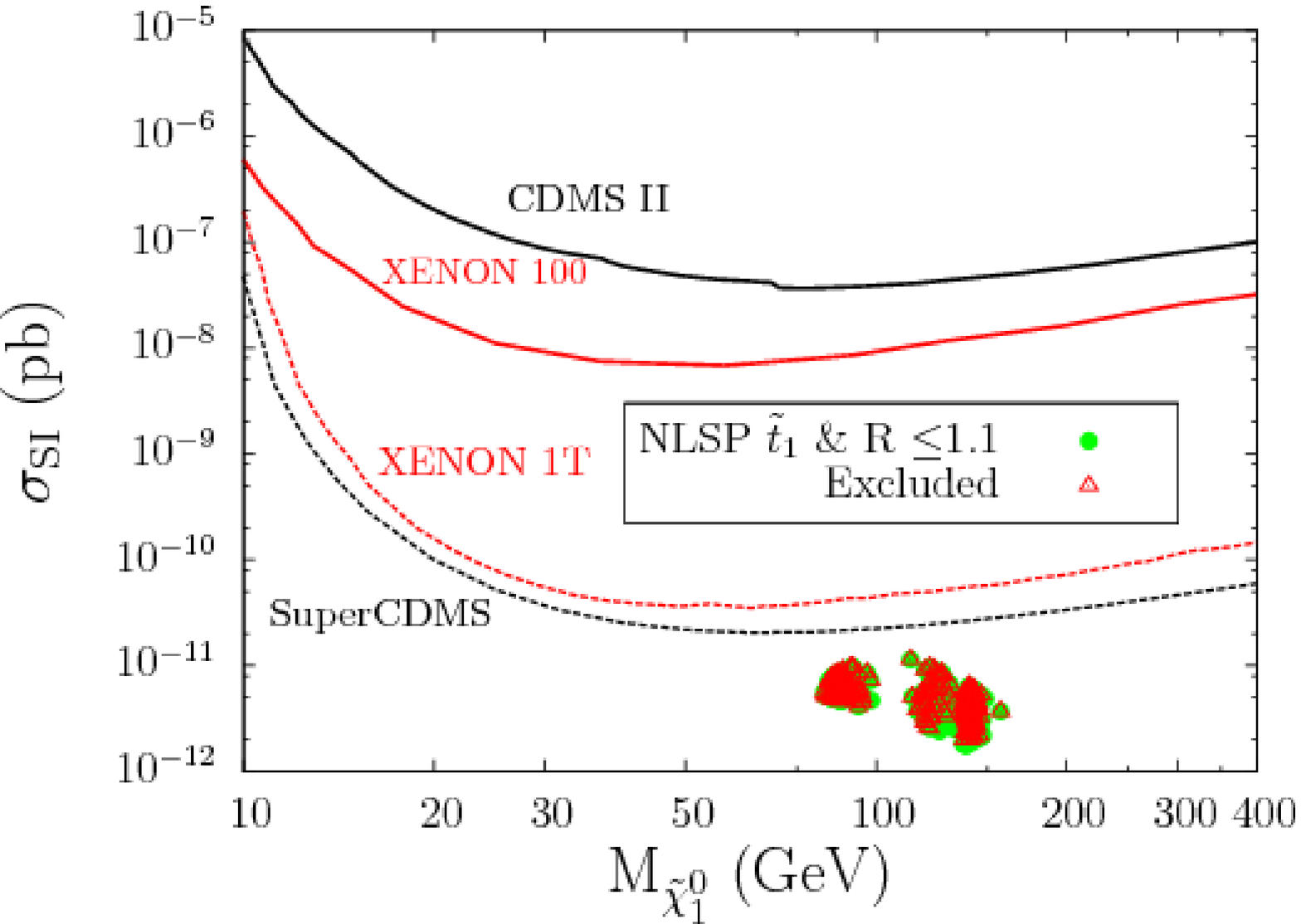}
\includegraphics[scale=1,width=8cm]{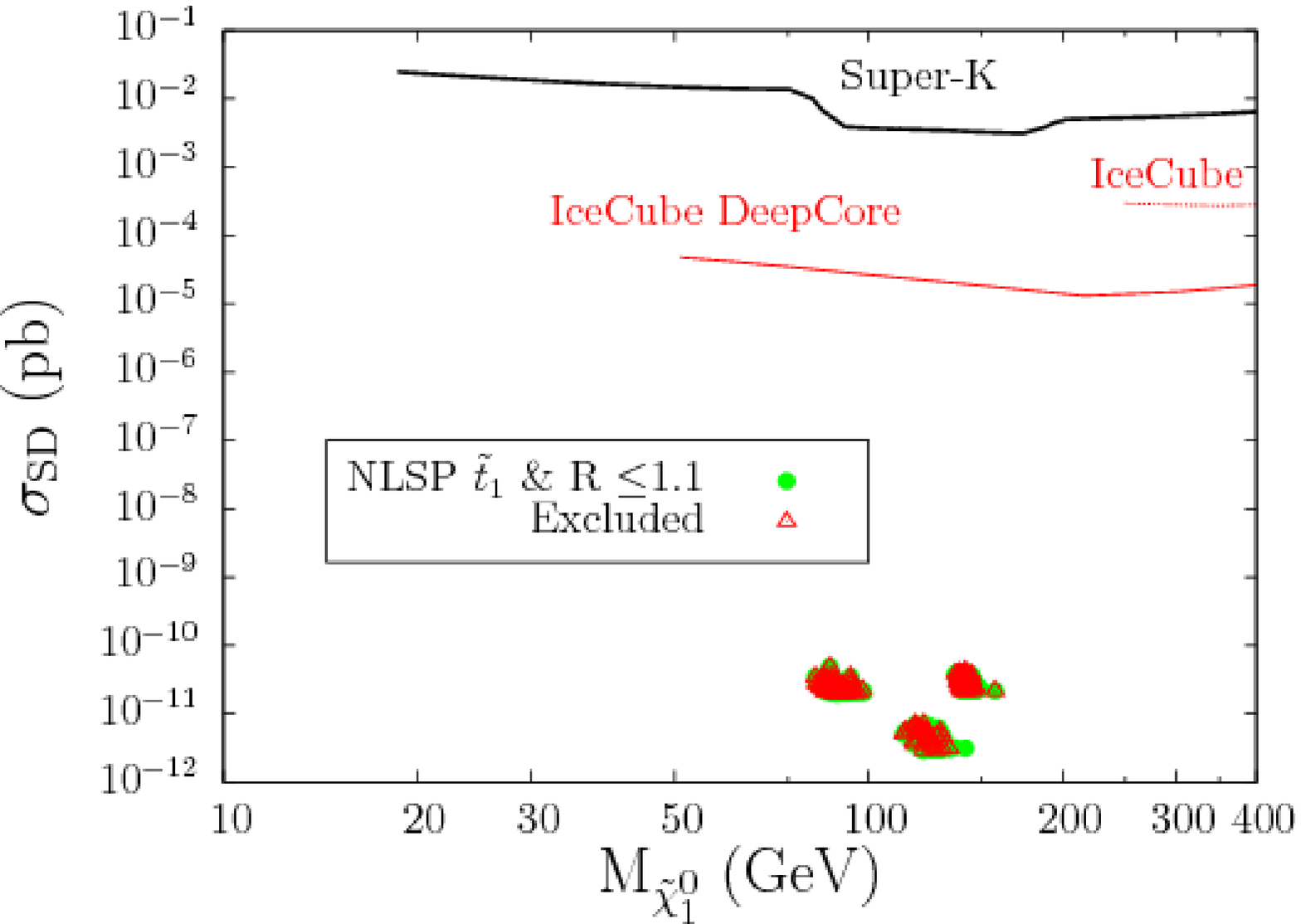}
\end{center}
\caption{$\sigma_{SI}$ (left panel) and $\sigma_{SD}$ (right panel) vs. $M_{\tilde{\chi}_1^0}$ in the framework of $b-\tau$ Yukawa unified mSUGRA/CMSSM. The excluded region is denoted in red. Current limits from CDMS-II, XENON100, SuperK and IceCube, and future projected sensitivities from XENON1T, SuperCDMS and IceCube DeepCore are also shown.} 
\label{dd}
\end{figure}

It is both interesting and important to see the implications of LHC data on direct and indirect dark matter detection in this class of Yukawa unified mSUGRA/CMSSM with NLSP stop. 
In Fig.~\ref{dd} we display this by plotting the spin-independent and spin-dependent WIMP-nucleon scattering cross section $\sigma_{SI}$ (left panel) and $\sigma_{SD}$ (right panel) vs. $M_{\tilde{\chi}_1^0}$. A significant region around $M_{\tilde{\chi}_1^0}\simeq 100$ GeV is excluded by LHC data, although it is allowed by CDMS-II, XENON100, SuperK and IceCube experiments. This excluded region even lies about one (six) order of magnitude below the expected XENON 1T/SuperCDMS (IceCube DeepCore) bound for spin-independent (spin-dependent) cross section.

\section{S\lowercase{ummary}}
Inspired by the recent LHC search for events containing monojet/multi-jets and large missing transverse momentum, corresponding to an integrated luminosity of 1 fb$^{-1}$, we have explored its ramifications for mSUGRA/CMSSM models which display $b-\tau$ Yukawa unification at 10\% level or better, contain NLSP stop, and possess LSP neutralino dark matter. In this Yukawa unification framework, the approximate mass degeneracy between NLSP stop and LSP neutralino require relatively large values of $m_0$ ($\sim 8-20$ TeV).
This coannihilation scenario with NLSP stop decaying into a soft jet evades the previous Tevatron bound. In terms of the emission of a hard QCD jet associated with stop pair production, followed by stop decay into a soft charm quark and LSP neutralino, we find that the monojet search at ATLAS is sensitive to the region with small mass difference between NLSP stop and LSP neutralino. The excluded limit can reach 160 GeV for the NLSP stop mass in the coannihilation region, while NLSP stop mass below 140 GeV is essentially excluded. The analysis for the production of stops based on the above searches excludes a significant parameter region in mSUGRA/CMSSM, namely in the region $8 \ {\rm TeV}\lesssim m_0\lesssim 16$ TeV.
The LHC implications for spin-dependent and spin-independent LSP neutralino-nucleon cross sections are also explored. Regions of the parameter space, some lying well below the much anticipated future bounds from IceCube DeepCore, Xenon 1T and SuperCDMS, are already excluded by utilizing the LHC data.

Note added: After this paper was essentially finished, several papers related to the NLSP stop scenario have appeared~\cite{wacker1,kats,sundrum,papucci,bi,in}.

\subsection*{Acknowledgment}
We would like to thank Ilia Gogoladze, Shabbar Raza, Daniel Feldman, Yevgeny Kats, Jing Shao and especially Alexander Belyaev for useful discussions. This work is supported by the DOE under grant No. DE-FG02-91ER40626. This work used the Extreme Science and Engineering Discovery Environment (XSEDE), which is supported by National Science Foundation grant number OCI-1053575.



\end{document}